\documentclass[useAMS,usenatbib,onecolumn,referee]{mn2e}
\usepackage{amssymb}
\usepackage{graphicx}
\usepackage{natbib}
\usepackage{bm}

\title[A possible explanation of the parallel tracks in kHz QPOs from LMXBs]{A possible explanation of the parallel tracks in kilohertz quasi-periodic oscillations from low-mass-X-ray binaries}
\author[Shi, Zhang \& Li]{ Chang-Sheng Shi$^{1,3}$\thanks{E-mail:
shics@hainu.edu.cn}, Shuang-Nan Zhang$^{2,5,6}$  and Xiang-Dong Li$^{3,4}$\\
$^{1}$College of Material Science and Chemical Engineering, Hainan
University, Hainan 570228, China\\
$^{2}$Key Laboratory of Particle Astrophysics, Institute of High Energy Physics, Chinese Academy of Sciences, Beijing 100049, China\\
$^{3}$Key Laboratory of Modern Astronomy and Astrophysics, Nanjing University, Ministry of Education, Nanjing 210093, China\\
$^{4}$Department of Astronomy, Nanjing University, Nanjing 210093, China;\\
$^{5}$National Astronomical Observatories, Chinese Academy of Sciences, Beijing 100012, China\\
$^{6}$University of Chinese Academy of Sciences, Beijing 100049, China\\}

\begin{document}

\date{Accepted ??; Received ??; in original form
 ??}

\pagerange{\pageref{firstpage}--\pageref{lastpage}} \pubyear{2017}

\maketitle

\label{firstpage}

\begin{abstract}
We recalculate the modes of the magnetohydrodynamics (MHD) waves in the MHD model (Shi, Zhang \& Li 2014) of the kilohertz quasi-periodic oscillations (kHz QPOs) in neutron star low mass X-ray binaries (NS-LMXBs), in which the compressed magnetosphere is considered. A method on point-by-point scanning for every parameter of a normal LMXBs is proposed to determine the wave number in a NS-LMXB.
Then dependence of the twin kHz QPO frequencies on accretion rates ($\dot{M}$) is obtained with the wave number and magnetic field ($B_{\ast}$) determined by our method. Based on
the MHD model, a new explanation of the parallel tracks, i.e. the slowly varying effective magnetic field leads to the shift of parallel tracks in a source, is presented. In this study, we obtain a simple power-law relation between the kHz QPO frequencies and $\dot{M}/B_{\ast}^2$ in those sources. Finally, we study the dependence of kHz quasi-periodic oscillation frequencies on the spin, mass and radius of a neutron star. We find that the effective magnetic field, the spin, mass and radius of a neutron star lead to the parallel tracks in different sources.
\end{abstract}

\begin{keywords}
accretion: accretion discs -- X-rays: binaries -- MHD -- stars: magnetic fields
\end{keywords}

\section{Introduction}
\label{intro}
The accretion process in an accretion disc has been studied for tens of years since the seminal work of Shakura \& Sunyaev (1973) and most observations in X-ray binaries (XBs) can be explained by accretion processes. Almost all the energy in our observation in XBs is released by accretion processes (Sunyaev \& Shakura 1986) and a kind of variability should been generated due to inner accretion disc oscillations in the boundary layer (Klu\'zniak \& Abramowicz 2005).

High frequency quasi-periodic oscillations (QPOs), have also been observed in NS-LMXBs, black hole LMXBs (see van der Klis et al. 1996, Strohmayer et al. 1996, Abramowicz \& Klu¡äzniak, 2001; van der Klis et al. 2006). A pair of kHz QPOs in NS-LMXBs often appear as twin peaks in their Fourier power spectra and are always labeled as the upper QPO and the lower QPO according to their frequencies ($\nu_{\rm u}$ \& $\nu_{\rm l}$; see Strohmayer et al. 1996; van der Klis 2000, 2006).
 Gilfanov et al. (2003) obtained the Fourier-resolved spectroscopy of 4U 1608$-$52 and GX 340+0 and suggested that the X-ray QPOs came from the boundary layer. Therefore millisecond variability is believed to originate from the innermost part of an accretion disc and is associated with strong gravity, and hence from which we might detect general relativistic effects.

To understand the physics of the kHz QPOs, many models have been proposed in order to explain these phenomena.
We can classify those models as four kinds: beat-frequency models (e.g. Miller et al. 1998), rotation, precession and epicyclic frequency models (e.g. Stella \& Vietri 1999), disc-oscillation and resonance models (e.g. Osherovich \& Titarchuk 1999;  Abramowicz \& Kluz\'niak W. 2001, 2003), wave models (e.g. Zhang 2004; Li \& Zhang 2005; Erkut et al. 2008; Shi \& Li 2009, 2010, Shi, Zhang \& Li, 2014). In most models, characteristic frequencies (e.g., the Keplerian frequency, the epicyclic frequency, Lense-Thirring precession frequency) in a geometrically thin accretion disc are always supposed as the central frequencies of kHz QPOs and are related with a characteristic radius.

The QPO frequencies in the power density spectra (PDS) always change with the source luminosity and inferred mass accretion rate (e.g., M\'endez et al. 1999; Belloni et al.
2005), but the relation between the kHz QPO frequencies in NS-LMXBs and their X-ray luminosity is puzzling. The correlation between kHz QPOs and X-ray luminosity is generally strong in a given source on a timescale of hours, but is absent both on longer timescales and among sources (see van der klis, 2003). Many authors (Zhang et al. 1998; Mendez \& van der Klis 1999; Mendez 2000; Van der Klis 2001; et al.) have discussed the interesting ``the parallel tracks" phenomenon that kHz QPOs and the X-ray intensity do not exhibit a one-to-one relation in 4U 1636$-$53 and 4U 1608$-$52. Van der klis (2001) pointed out that the phenomenon is particularly evident in kHz QPO sources (the near-Eddington neutron star Z sources and the less luminous neutron star atoll sources), and proposed that at least two time-variable and independent
parameters are involved.

Erkut et al. (2016) studied the correlation between the kHz QPO frequency and X-ray luminosity or  $\dot{M}/B_{\rm NS}^2$ in an ensemble of NS-LMXBs, where $\dot{M}$ is an accretion rate in NS-LMXBs and $B_{\rm NS}$ is the surface magnetic field strength of a neutron star. Then they found a correlation described by a power-law function between $\nu_{\rm l}$ and $\dot{M}/B_{\rm NS}^2$. They suggested that $\dot{M},\ \dot{M}/B_{\rm NS}^2$ are expected to determine the QPO frequencies. The interdependence between $\nu_{\rm l}$ and $\dot{M},\ \dot{M}/B_{\rm NS}^2$ in the ensemble of NS-LMXBs is a new avenue to explore the correlation between the kHz QPO frequency and X-ray luminosity.
Erkut \& \c{C}atmabacak (2017) explained the parallel tracks using the boundary region model of kHz QPOs (Alpar \& Psaltis 2008; Erkut, Psaltis, \& Alpar 2008).
In that model, the properties of the innermost disk region (e.g. the radial width of the
boundary layer, the magnetospheric radius, and the magnetic
diffusivity) vary with $\dot{M}$ and determine the kHz QPO frequencies.

 Shi \& Li (2009, 2010) obtained two MHD oscillation modes in NS-LMXBs and BH-LMXBs respectively, which are supposed as the sources of high frequency QPOs. Shi, Zhang \& Li (2014) also considered the two MHD oscillation modes at the magnetosphere radius between a NS magnetosphere and a standard thin accretion disc as the origin of kHz QPOs, but the compression effect of the magnetosphere is considered. In our model, the relation between the frequencies of the twin peak kHz-QPOs and the accretion rate is obtained.  Recently simulations on the accretion process were performed in order to find the nature of QPOs. A three-dimensional MHD simulation of an accretion process at a quasi-equilibrium state was performed when the gravitational, centrifugal, and pressure gradient force were considered by Kulkarni \& Romanova (2013). Parthasarathy et al. (2017) also performed an idealised MHD simulation of oscillating cusp-filling tori orbiting a non-rotating axisymmetric NS and they confirmed the modulation mechanism of the neutron star boundary layer luminosity by disc oscillations.

In this work, we recalculate the modes of the MHD waves in the MHD model of kHz QPOs in NS-LMXBs (Shi, Zhang \& Li 2014) with a new method of determining several parameters. The explanation of the parallel tracks, in which the slowly varying effective magnetic fields lead to the parallel tracks in a source, is presented. We start in section 2 with the review of
the MHD model (Shi, Zhang \& Li 2014). In section 3 we study the dependence of kHz QPO frequencies on several parameters and suggest a new explanation of the parallel tracks. In section 4 we make a discussion on the effective magnetic field and the physics of QPOs. Finally, we summarize our result in section 5.

\section{REVIEW OF OUR MHD MODEL}
In our MHD model (Shi, Zhang \& Li 2014), the stellar magnetosphere is compressed by the accretion matter in a steady standard $\alpha$-disc of Shakura \& Sunyaev (1973) in a NS-LMXB, i.e. the disc is truncated by the stellar magnetosphere. As shown in Figure 1, the accretion flows pass through a narrow boundary layer by an exchange instability (Spruit and Taam 1990, Kulkarni \& Romanova 2013).
The plasma is eventually accreted to the polar cap along the magnetic field lines (Ghosh et al. 1977, Gosh \& Lamb 1979). In the
accretion process, the accreted plasma hits the magnetic field lines and a deformation of the magnetic field emerges. Then the balance at the magnetosphere radius from the competition of gravity, inertial centrifugal force,
Coriolis force, magnetic pressure and gas pressure is interrupted due to a small turbulence and some instability may arise (Elsner \& Lamb 1977). The MHD waves generated at the magnetosphere radius modulate the luminosity and lead to the kHz QPOs (e.g. Parthasarathy et al., 2017).
\begin{figure}[h]
\begin{center}
\includegraphics[width=0.5\columnwidth]{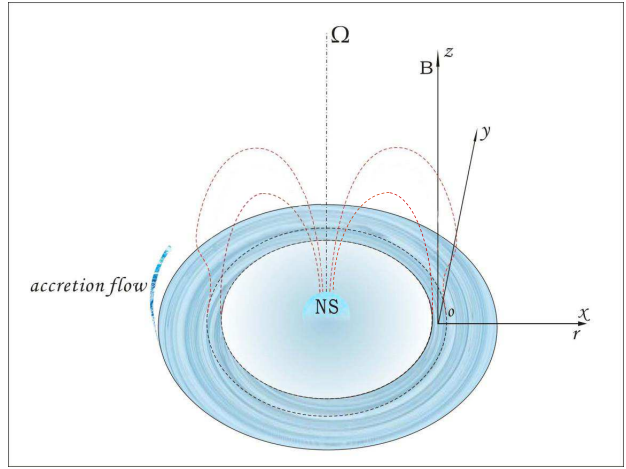}
  \caption{A sketch of the accretion process in NS-LMXBs when the compressed magnetosphere is considered.} \label{fig1}
 \end{center}
\end{figure}

\subsection{Modes of MHD Waves}
In a NS-LMXB, the plasma from the companion is transferred into
the magnetosphere radius by the accretion disc and then it is in the equilibrium state.
In the equilibrium state, the balance equation for the plasma at the magnetosphere radius can be expressed as follows (Shi, Zhang \& Li 2014):
\begin{equation}
\label{eq1}
\rho_0 \frac{\partial \bm u_0 }{\partial t} + \rho _0 \bm u_0 \cdot
\nabla \bm u_0 = - \rho _0 \textstyle{{GM} \over {r_0
^3}}\bm r_0+ \rho _0 \Omega ^2 \bm r_0 -
\rho _0 (\bm \Omega \cdot \bm r_0 )\bm \Omega + 2\rho _0 \bm u_0 \times \bm \Omega + \textstyle{1 \over \mu_0 }(\nabla \times
\bm B_0 )\times \bm B_0- \nabla P_0,
\end{equation}
where $\rho$ is the plasma density, ${\bm u}$ the plasma velocity, $G$ the gravitational constant, $M$ the mass of the NS, $\bm r$ the displacement from the NS, $P$ the barometric pressure, $\mu_0$ the vacuum magnetic conductivity, $\bm{B}$ the magnetic field at the magnetosphere radius, and ${\bm \Omega}$ the angular velocity of the NS respectively. We label the variables in the equilibrium state with the subscript ``0" and the variation
of a physical quantity due to the perturbation with a subscript ``s".  In this study the letters in bold italic also express vectors.

In the model, we consider MHD equations in a co-rotation frame of reference and adopt several conditions to obtain the MHD modes of the MHD waves.
(1) The balance is supposed to be steady at first, i.e.,
$ \frac{\partial P_0 }{\partial t} = 0$, $\frac{\partial \rho _0 }{\partial t}=0$,
$\frac{\partial { {\bm B}}_{\rm {\bf 0}} }{\partial t} = \bm 0$ and $\frac{\partial
{ {\bm \Omega }}}{\partial t} = \bm 0$. (2) Next, a small disturbance by strong excitation (Rezania \& Samson 2005) leads to the MHD waves. So the the variation
of a physical quantity due to the small perturbation is much smaller than the variable in balance, $u_{\rm s} \ll \left| {\bm{\Omega} \times \bm{r}} \right|$, $B_{\rm s} \ll B_0$, $r_{\rm s} \ll r_0$, $\rho_{\rm s} \ll \rho_0$, $P_{\rm s} \ll P_0$.  (3) The plasma is an ideal conductor, i.e. the vacuum electroconductibility $\sigma\rightarrow \infty$. (4) The accretion disc does not warp, i.e. the rotation axis of the accretion disc is parallel to the spin of the NS. Then we can obtain the spin $\bm{\Omega} = (0, 0, \Omega)$ and that the uncompressed dipole magnetic field lines or the compressed magnetic field lines are normal to the disc in the equatorial plane. (5) The density and magnetic field are only functions of the longitudinal displacement ($r_0$) after being compressed, i.e. $\rho_0 = \rho_0(r_0)$, $\bm{B_0} = (0, 0, B_0(r_0))$. (6) The MHD waves propagate along the magnetic field lines or they would be dissipated in the disc easily (Shi \& Li 2009), i.e. the wave vector can be expressed as, $\bm k = (0,0,k)$, here $k$ is the wave number.

Now the balance equation of the plasma at the magnetosphere radius can be simplified as
follows (see Shi, Zhang \& Li 2014),
\begin{equation}
\label{eq2}
\Omega ^2r_0 - \textstyle{{GM} \over {r_0 ^3}}r_0 = (\frac{c_{\rm s}^2}{\rho _0
}\frac{\partial \rho _0 }{\partial r} +
\frac{1}{\mu _0 \rho _0 }B_0 \frac{\partial B_0 }{\partial r})\mid_{r=r_0},
\end{equation}
where ${c_{\rm s} = \sqrt{\frac{\gamma P_0}{\rho_0}}}$ is the sound velocity and the adiabatic index $\gamma = \frac{5}{3}$ is adopted.

The equation according to the Faraday
principle of electromagnet induction, the continuity and the adiabatic condition can be expressed as follows,
\begin{equation}
\label{eq3} \frac{\partial { {\bm B}}_{ { 0}} }{\partial t} = \nabla \times
({ {\bm u}_{0}}\times {\bm B}_{ {0}}),
\end{equation}

\begin{equation}
\label{eq4} {\partial{\rho_0} \over \partial t}
+\nabla\cdot({\rho_0}{{\bm u_0}})=0,
\end{equation}

\begin{equation}
\label{eq5} P_0\rho_0^{-\gamma}={\rm const}.
\end{equation}

According to the above equations in the equilibrium state, we can obtain the full equations with the same expressions for the plasma in a state after being disturbed due to some turbulance or some instability by changing
$\rho_0, P_0, u_0, {\bm B}_{ {0}}, \bm r_0$  with $\rho_0 + \rho_{\rm s}$, $P_0 + P_{\rm s}$, $u_0+u_{\rm s}$, ${\bm B}_{ {0}}+{\bm B}_{\rm s}$, ${\bm r}_0 + {\bm r}_{\rm s}$. Then we can obtain the equations about the perturbation quantities with the above two groups of equations (see Shi, Zhang \& Li 2014). After carrying out Fourier transformation ($f \rightarrow f e^{i {\bm k \cdot \bm r} - i \omega t}$) for the equations about the variables due to a small perturbation ($u_{\rm s}$, $B_{\rm s}$, $r_{\rm s}$, $\rho_{\rm s}$, $P_{\rm s}$), we can obtain the last dispersion equation:
\begin{equation}
\label{6} \begin{array}{lll}
( - \omega ^2 - 2\Omega ^2 - 4\omega _{\rm k}^2 \mbox{ + }k^2V_{\rm A}^2)(- \omega ^2 - \Omega ^2 + \omega _{\rm k}^2 \mbox{ + }k^2V_{\rm A}^2)(- \omega ^2 + \omega _{\rm k}^2 \mbox{ + }k^2c_{\rm s}^2)
-4 \Omega ^2 \omega^2 (- \omega ^2 + \omega _{\rm k}^2 \mbox{ + }k^2c_{\rm s}^2)\\
=k^2 r_0^2 (\Omega ^2 - \omega _{\rm k}^2)^2 (- \omega ^2 - \Omega ^2 + \omega _{\rm k}^2 \mbox{ + }k^2V_{\rm A}^2) - (\Omega ^2 - \omega _{\rm k}^2) (- \omega ^2 - \Omega ^2 + \omega _{\rm k}^2 \mbox{ + }k^2V_{\rm A}^2)\gamma k^2 c_{\rm s}^2 \frac{r_0}{\rho_0} (\frac{\partial \rho_0}{\partial r}\mid_{r=r_0}),
\end{array}
\end{equation}
where ${\omega_{\rm k} = \sqrt{\frac{GM }{r_0^3}}}$ is the Kepler angular frequency, ${V_{\rm A} = \sqrt{\frac{B_0^2}{{\mu \rho_0}}}}$ the Alfv\'en velocity and ${c_{\rm s} = \sqrt{\frac{\gamma P}{{\rho_0}}}}$ the sonic velocity respectively.

The frequency solutions ($\nu={\omega}/2\pi$) of Equation (6) are supposed as the frequencies of the kHz QPOs in NS-LMXBs. Analytic solutions of Equation (6) is too long to be used and thus numerical solutions are found for a given magnetosphere radius.

\subsection{The Magnetosphere Radius and Compression Process}

The spherical accretion process onto a compact star was discussed by Lamb et al. (1973)
at first and they gave a definition of the magnetosphere
radius, in which the plasma is in a balance under the interaction of all the forces. Then many authors (e.g., Elsner \& Lamb 1977; Burnard
et al. 1983; Mitra 1992; Li \& Wang 1995; Weng \&
Zhang 2011) obtained a similar relation between the outer boundary of the
magnetosphere in a pulsar
and the Alfv\'en radius, i.e. $ r_{\rm ms} = c1*r_{\rm A}\ $ ($c1$ is a coefficient), where the Alfv\'en radius can be expressed as follows (Lamb et al. 1973),
\begin{equation}
\label{eq7}
r_{\rm A} \simeq 2.29\ast 10^6 ({\dot{M}_{16}}/{B_{\rm NS, 8}^2})^{-2/7} M_{1.4\odot}^{-1/7} R_{6}^{12/7}\ {\rm cm},
\end{equation}
where $B_{\rm NS}$ is the surface magnetic field of a NS, $R$ is the radius of a NS and the subscripts ``8", ``16", ``$1.4\odot$", ``$6$" express the quantities in units of $10^{8}\ { \rm G}$, $10^{16}\ {\rm g/s}$, 1.4 times the mass of the Sun, $10^{6}\ { \rm cm }$, respectively.

Kulkarni \& Romanova (2013) considered a quasi-equilibrium state, in which the pressure gradient forces, the gravity and the centrifugal force in a compressed non-dipolar magnetic field are in balance and made a three-dimensional MHD simulations of the magnetospheric accretion. Then they obtained a different $r_{\rm m}$, which depends on the magnetic moment, mass, radius of the NS and $\dot{M}$, as follows,
\begin{equation}
\label{eq8}
r_{\rm m1} \approx {2.50}\times 10^6 \mu_{26}^{2/5} \dot{M}_{16}^{-
1/5} m_{1.4\odot}^{-1/10} R_6^{3/10}\ {\rm cm},
\end{equation}
where $\mu$ is the effective magnetic moment of a NS, which can be significantly different from a dipolar moment, and the subscript ``$26$'' expresses the quantity in unit of $10^{26}\ {\rm G\cdot m^3}$.

As discussed by Arons (1993), the magnetic moment $\mu$ in non-dipolar magnetic field is not $B_{\rm NS}R^3$. We thus parameterize $\mu$ for the compressed magnetic field as $B_{\rm \ast}R^3=c B_{\rm NS}R^3$, where $c$ is a simply form factor that depends on the shape of the magnetic field and the angle between the spin and the magnetic axis, and $B_{\rm \ast}$ is the effective surface magnetic field of a NS. The magnetic field form factor ($c$) will be discussed in section 4. After the magnetic moment is substituted, Equation 8 can be written as,
\begin{equation}
\label{eq9}
r_{\rm m1} \simeq  {2.50} \times 10^6 ({\dot{M}_{16}}/{B_{\ast 8}^2})^{-1/5} M_{1.4\odot}^{-1/10} R_6^{3/2}\ {\rm cm}.
 \end{equation}

Shi, Zhang \& Li (2014, 2015) also estimated the magnetosphere radius by the balance of the plasma in the presence of the magnetic pressure, the barometric pressure and the collision ( ${\frac{B^2}{8 \pi} \simeq P + \rho u^2}$, Eslner \& lamb 1977; Romanova et al. 2002) for the dipolar magnetic field and the standard thin disc in NS-LMXBs.
The magnetosphere radius can be expressed as,
\begin{equation}
\label{eq10}
r_{\rm m2} \simeq 3.59 \times 10^5 \alpha_{0.1}^{4 / 15}
({\dot{M}_{16}}/{B_{\rm NS, 8}^2})^{ - 34 / 135} B_{\ast 8}^{4 / 45} M_{1.4\odot}^{ - 7 / 27} R_{6}^{16 / 9} f^{ - 136 / 135}\ {\rm cm},
\end{equation}
where ${f = (\mbox{1} - \sqrt {\frac{R}{r}} )^{1 / 4}}$, $\alpha$ is the viscosity parameter (Frank et al. 2002).

$r_{\rm m1}$ and $r_{\rm m2}$ are applied into Equation (6) and correspond to two models: model-1 ($r_{\rm m1}$) with a compressed magnetic field and a standard thin accretion disc but model-2 ($r_{\rm m2}$) with the dipolar magnetic field and a compressed accretion disc. Model-2 needs an appropriate density of the compressed disc that is however not available, and so we use both the dipolar magnetic field and a density of a standard thin accretion disc in order to estimate the QPO frequencies. In model-1, the density of the standard thin accretion disc and a compressed magnetic field from the conservation of the magnetic flux are substituted and the result fits the observation with our method of determining $\dot{M}$ better than model-2 (see Shi, Zhang \& Li). $r_{\rm m1}$ is the characteristic radius of the kHz QPOs in the models and we will only consider model-1 below. With $r_{\rm m1}$, i.e. the radius in the impact balance, the relevant physical quantities for Equation (6) can be expressed as follows,
 \begin{equation}
\label{eq11}
\omega_{\rm k} = {3442.06} ({\dot{M}_{16}}/{B_{\ast 8}^2})^{3/10} M_{1.4 \odot}^{13/20} R_{6}^{-9/4},
\end{equation}
 \begin{equation}
\label{eq12}
c_{\rm s} = {4.08} \times 10^7 ({\dot{M}_{16}}/{B_{\ast 8}^2})^{9/40} B_{\ast8}^{3/10} M_{1.4 \odot}^{13/80} f^{3/5},
\end{equation}
 \begin{equation}
\label{eq13}
V_{\rm A}  = {5.41 \ast 10^7} ({\dot{M}_{16}}/{B_{\ast 8}^2})^{-37/80} B_{\ast8}^{9/20} (r_{\rm m1,6}^2-r_{\rm m2,6}^2)^{ - 1}r_{\rm m2,6}^{-1} f^{-11/10}\ {\rm cm / s},
\end{equation}
\begin{equation}
\label{eq14}
\frac{r_0}{\rho_0} \frac{\partial \rho_0}{\partial r}\mid_{r=r_0} = 4.55 \ast \alpha _{0.1}^{ - 7 / 10} M_{1.4 \odot}^{29 / 40} R_6^{ -
3 / 2} [ - 3.68 + 0.14({\dot{M}_{16}}/{B_{\ast 8}^2})^{1 / 10}R_6^{ - 3 / 4} M_{1.4 \odot}^{1 / 20} f^{ - 4}].
\end{equation}
According to Equations (6),(9)-(14) and the expression of $f$, we can obtain the relation between $\nu_{\rm l},\ \nu_{\rm u}$ and $\dot{M}$. As discussed in Shi, Zhang \& Li (2014), the Kepler angular frequency is higher than $k V_{\rm A}$ and $k c_{\rm s}$, and is the key factor to determine the solutions of Equation (6). $\omega_{\rm k}$ has often been adopted as $\nu_{\rm u}$ in other researches, which might be a very crude approximation without considering the spin of a NS.
The quantity ($k^2 v_{\rm A}^2+\omega_{\rm k}^2-\Omega^2$) decreases from a positive value into a negative value with the decrease of $\dot{M}$ and meanwhile the twin solutions change into a single solution, i.e. the quantity determines the number of the real solutions of Equation (6) (i.e. the frequencies of QPOs in NS-LMXBs). As seen from the above equations, it is obvious that the physical variables are dependent on ${\dot{M}_{16}}/{B_{\ast 8}^2}$, which will be discussed below.

\section{PARAMETERS DEPENDENCE}
Since the detection of QPOs in LMXBs for about thirty years, the phenomena in a single source and an ensemble of many sources have been explored. Some parameters (e.g. the spin, mass and radius of a NS) from a correct QPO model  are a good index used to restrict the equation of state (EOS) of a NS.
As described by van der klis (2006), a good correlation between kHz QPOs and X-ray luminosity can keep for several hours but be absent both on longer timescales and between sources. It is obvious that not one but several parameters control the behaviors of QPOs. Therefore the dependence of the solutions on the related parameters in our MHD model should be explored.

\subsection{The MHD Wave Number}
The wave number ($k$) of the MHD wave driven by a
disturbance in the magnetosphere radius is mainly determined
by the size of a NS (Rezania \& Samson 2005).
Shi, Zhang \& Li (2014) selected $k$ by comparing the calculated results to the observation and the best one that matches the most data was determined by eye. To improve the method, we propose a method in order to select the best $k$ to match the observation of the central frequencies of the twin kHz QPOs.

Similar to Shi, Zhang \& Li (2014), the key point to select the best $k$ is comparing the calculated results to the observation. Because $\dot{M}$ in most NS-LMXBs is not known, $\dot{M}$ can be determined by $\nu_{\rm lo}$ or $\nu_{\rm uo}$, where $\nu_{\rm lo}$ is the observed frequency of a lower QPO and $\nu_{\rm uo}$ is the observed frequency of an upper QPO. The modes of the MHD waves are determined by Equation (6), so one of the twin solutions exhibit a one-to-one relation with the other one. The solutions of Equation (6) vary monotonically with increasing $\dot{M}$, so we can select a group of $\dot{M}, \nu_{\rm lc}, \nu_{\rm uc}$ according to the comparison criteria ($\nu_{\rm lo}=\nu_{\rm lc}$) in theory, where $\nu_{\rm lc}, \nu_{\rm uc}$ are the lower and upper real solutions of Equation (6). Then we can compare $\nu_{\rm uo}$ with $\nu_{\rm uc}$ and test our MHD model.
However the data of $\nu_{\rm lc}, \nu_{\rm uc}$ are not continuous due to the discontinuity of $\dot{M}$ adopted in our numerical calculation, the comparison criteria is taken as $\mid\nu_{\rm lo}-\nu_{\rm lc}\mid \lesssim 3\ {\rm Hz}$ below.

 Here we consider the standard parameters of a NS, i.e.,
$M = 1.4 M_{\odot}$, $R = 10^6\ {\rm cm}$, $\alpha = 0.1$ (King et al. 2007) and calculate the minimum deviation to select the value of $k$
.
\begin{enumerate}
\item Choose the best initial value of $B_{*}$ ( e.g. the result in Shi, Zhang \& Li, 2014 or the characteristic value $B_{*}=10^8\ {\rm G}$ for LMXBs) and $k$ for a NS with known spin.

\item Substitute one $\dot{M}$ into $r_{\rm m1}$  and Equation (11)-(14).

\item Solve Equation (6), the frequency of a single solution ($\nu_{\rm sc}$) or $\nu_{\rm lc}, \nu_{\rm uc}$ are obtained.

\item Select different $\dot{M}$ (e.g. $\dot{M}$ changed in a small step size in this study) and repeat step (2) and (3), then one possible relation between $\nu_{\rm sc}$ or $\nu_{\rm lc}, \nu_{\rm uc}$ and $\dot{M}$ is obtained.

 \item Determine $\dot{M}$ by $\nu_{\rm lc}$ when $|\nu_{\rm lo}-\nu_{\rm lc}| < 3\ {\rm Hz}$.
     Then one or more groups of values of $\dot{M},\nu_{\rm lc},\nu_{\rm uc}$ corresponding to one group of values of $\nu_{\rm lo},\nu_{\rm uo}$ are selected.

 \item Select all $\nu_{\rm uc}$ that match the condition in step (5) for all $\nu_{\rm lo}, \nu_{\rm uo}$ in a source and count their numbers as $n$. Then
     calculate the average of the
     the sum of deviation square $\sigma = \frac{1}{n} \sqrt{\Sigma (\nu_{\rm uc} - \nu_{\rm uo})^2}$.

 \item Calculate all $\sigma$ for different $k$ and then select $k$ with the minimum $\sigma$.
\end{enumerate}

In step (5) we require that the maximum difference between $\nu_{\rm lc}$ and $\nu_{\rm lo}$ as around $3\ {\rm Hz}$, because the errors of the central frequencies of most observed twin kHz QPOs are larger than $3\ {\rm Hz}$ (see the data from van Straaten et al. 2000, 2002, Altamirano et al. 2008, Di Salvo, M\'endez
et \& van der Klis 2003, Jonker, M\'endez \& van der Klis 2002, Wijands et al. 1997, van Straaten, van der Klis \& M'endez 2003).

As shown in the left panel of Figure 2, $k$ corresponding to the smallest $\sigma$ can be obtained and selected to describe the kHz QPOs in one NS-LMXB. The selected best $k$ is used for the relation
between $B_{\ast}$ and $\sigma$ in the right panel of Figure 2. We obtain the average values of $\sigma$ for 4U 1608$-$52, 4U 0614+09, 4U 1636$-$53, which is labeled as $\langle\sigma\rangle$, are $26.98, 33.02, 36.10\ {\rm Hz}$, respectively. Due to the maximum difference value ($3\ {\rm Hz}$) of $\nu_{\rm lc}$ for determining $\dot{M}$ in step (5), the maximum difference value ($|\sigma-\langle\sigma\rangle|$) may also reach up to $3\ {\rm Hz}$ as shown in the right panel, if the step size of selected $\dot{M}$ is small enough. According to our calculation, $\sigma$ is closer to $\langle\sigma\rangle$ if the step size of selected $\dot{M}$ is smaller, i.e. $\sigma$ can be thought as an invariant with the increasing value of $B_{\ast}$. The three values of $\langle\sigma\rangle$ among 4U 1608$-$52, 4U 0614+09, 4U 1636$-$53 are different, which may be caused by the other undetermined physical quantities such as $M,\ R$ for different sources.

The above result means that $B_{*}$ for an appropriate range (e.g. a normal magnetic field of a NS in our calculation) can not affect the selection of $k$. Therefore the best $B_{*}$ also can not be selected by $\sigma$, but we can select it according to the observation with $\dot{M}$ (e.g. Ford, et al. 2000) for two criteria: (1) the selected $\dot{M}$ of the kHz QPOs in step (5) should be in the range of $\dot{M}$ in NS-LMXBs from observation, i.e. the data should match the observation as much as possible; (2) $\sigma$ should be as small as possible when condition (1) is satisfied. This selection method for the derived $B_{*}$ has a small randomness
due to the almost constant $\sigma$ for different $B_{*}$.

\begin{center}
\begin{figure*}[h]
\label{fig2}
 \includegraphics[width=1\columnwidth]{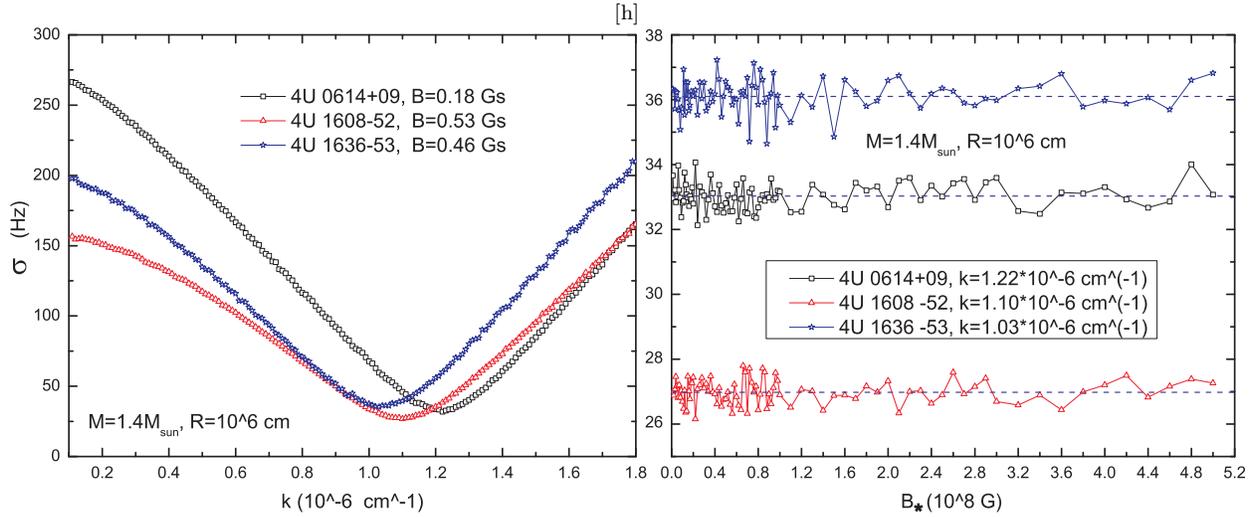}
 \caption{ {\it Left}: Relation between $k$ and $\sigma$; {\it right}: relation between $B_{\ast}$ and $\sigma$.}
\end{figure*}
\end{center}

\subsection{Comparison with Observations}
 According to the coherence on the frequencies of QPOs in two different frames (Shi, 2011), we can compare the numerical solutions with the observed kHz QPOs. Considering the different $\dot{M}$ and the characteristic parameters of NS-LMXBs, i.e., $m = 1.4 M_\odot$, $R = 10^6\ {\rm cm}$, and $\alpha = 0.1$ (King et al. 2007), we recalculate the numerical solutions of Equation (6) with the new parameter values, $B_*, k$, which is obtained by the method in Subsection 3.1 and is shown in Table 1.
The transition radius ($r_{\rm t}$) between the single solution and the twin solutions for the MHD model is very near the corotation radius, which means that the locations of almost all twin kHz QPOs lie inside the corotation radii.

\begin{table*}
 \centering
  \caption{This table lists the selected parameters ($k, B_{\ast}$) in our numerical calculations, the known $\nu_{\rm NS} = \frac{\Omega}{2\pi}$, the derived corotation radius ($r_{\rm co}$) and the derived transition radius ($r_{\rm t}$) between $\nu_{\rm sc}$ and $\nu_{\rm lc}, \nu_{\rm uc}$ from our MHD model. } \label{t:1}
  \begin{tabular}{@{}lccccc@{}}
   \hline
  \hline
 {sources} & $k$ & $B_{\ast}$ & $\nu_{\rm NS}$ & $r_{\rm co}$ & $r_{\rm t}$\\
   & $ (10^{-6}\ {\rm cm^{-1}})$ & $ (10^8\ {\rm G})$& $({\rm Hz})$ &$ (10^6\ {\rm cm})$ & $ (10^6\ {\rm cm})$ \\
 \hline
4U 0614+09 & 1.22 & 0.55 & 415  & 3.011694  & 3.011707\\

4U 1608$-$52 & 1.10 & 0.64 & 619  & 2.307009  & 2.307013\\

4U 1636$-$53 & 1.03 & 1.24 & 581  & 2.406535  & 2.406543\\
\hline
\end{tabular}
\end{table*}

\begin{figure*}[!htbp]
\begin{center}
\label{fig3}
\includegraphics[width=0.6\columnwidth]{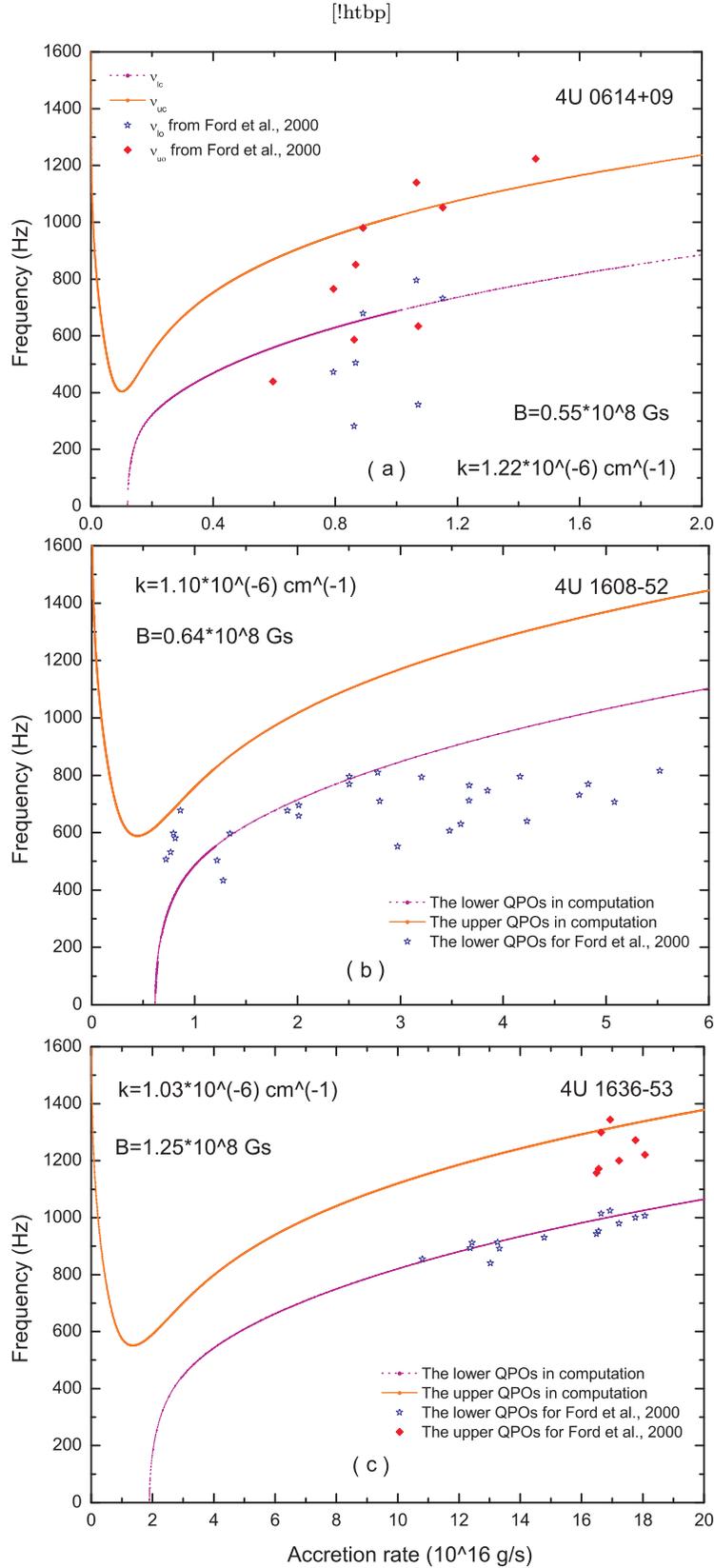}
\caption{ Relations between $\nu_{\rm u}$, $\nu_{\rm l}$, and $\dot{M}$ for
three sources (for the measured data: Ford, 2000). The solid lines express $\nu_{\rm s}$ or $\nu_{\rm u}$ and the dotted lines express $\nu_{\rm l}$ from our calculation. Note: the dots in the dotted lines are not uniformly distributed due to the nonlinear change of $\nu_{\rm u}$, $\nu_{\rm l}$ with the increasing $\dot{M}$ adopted the same step size in step (4), as is the same in the following figures.}
\end{center}
\end{figure*}

As shown in Figure 3, the relations between $\nu_{\rm u}$, $\nu_{\rm l}$ and $\dot{M}$ are obtained when $k,\ B_{\ast}$ are selected. The solid lines in Figure 3 come from our numerical solutions for $\nu_{\rm u}$ and the dotted lines for $\nu_{\rm l}$ from our calculation. The data (Ford et al., 2000) are scattered around the lines and $\dot{M}$ corresponding to the data from Ford et al. (2000) is obtained by $L=GM\dot{M}/R$.
The data from Ford et al. (2000) scatter in Figure 3, which may correspond to the parallel tracks phenomenon.

Van der Klis (2001) adopted a composite luminosity, $L_x \propto \dot{M}(t)+c_2 \langle\dot{M}\rangle(t)$, in order to explain the parallel tracks, where $L_x$ is the observed X-ray luminosity, $c_2$ a constant, $\dot{M}(t)$ the instantaneous
accretion rate through the disc and $\langle\dot{M}\rangle(t)$ the running average of accretion rate. The instantaneous
accretion rate is a fast varying component and the running average is a slowly varying component. The relation between the frequencies of kHz QPOs and accretion rates was supposed as $\nu(t)\propto [{\dot{M}(t)/L_x(t)}]^\beta$, where $\beta$ was a positive parameter and $\nu(t)$ was a kHz QPO frequency as a function of time, but its physical mechanism need to be explored. In addition, van der Klis (2001) also proposed that magnetic-field strength related to average $L_x$ might explain that different sources have different tracks. We suggest an explanation of the parallel tracks when slowly variable effective magnetic fields are considered.

\subsection{A New Explanation of the Parallel Tracks}
The parallel tracks phenomenon has been reported by M\'endez et al. (1999), Ford et al. (2000), van der Klis (2001) and so on. As seen from their observations, we can find that the parallel tracks do not spread continuously in a frequency-accretion rate plane, i.e. the tracks seems to be shifted between different lines. It means that the accretion system might be shifted from one state to another state and one or more variable quantities modulate the shift of different tracks. Of course the appearance of discrete tracks can also an observational effect due to discontinuous sampling and the infrequent visibility of QPOs, which needs more observation and will not be discussed here.

The same as Van der Klis (2001), we also consider that the inner radius $r_{\rm m}$ of the disc, and hence $\nu$, is determined by a balance, which comes from the competition of gravity, inertial centrifugal force, Coriolis force, magnetic pressure and gas pressure.
In a given source on a timescale of hours, the frequency ($\nu$) of the MHD wave originating from a small disturbance is mainly determined by $\dot{M}$. But on longer timescales, the magnetic field near the magnetosphere radius might be changed due to the deformation of the magnetic field by some instabilities and some violent activity such as magnetic reconnection. Because $B_{\ast}$ is obtained by the definition of magnetosphere radius and the hypothesis about the shape of the magnetic field (e.g. the dipolar magnetic field: $B_{\rm NS}R^3/r^3$), the changed
magnetic field near the magnetosphere radius leads to change of $B_{\ast}$, which is discussed in subsection 4.
Thus the track of kHz QPOs will shift in the frequency-accretion rate plane. Between different sources, their $B_{\ast},\ \nu_{\rm NS},\ M,\ R$ may not be the same, so their tracks of kHz QPOs do not overlap with each other in the frequency-accretion rate plane. Both on longer timescales and between sources, we suppose that the physics of the kHz QPOs is the same, i.e. the lines in Figure 3 and 4 should be similar. The instantaneous
accretion rate is a fast varying component and the derived $B_{\ast}$ is a slowly varying component. The effective magnetic field $B_{\ast}$ modulate the changing of the lines (i.e. the tracks).

Erkut et al. (2016) studied the dependence of $\nu_{\rm l}$ on $\dot{M}$ and $\dot{M}/B_{\ast}^2$ in the ensemble of neutron star low-mass X-ray binaries.
The ensemble trend in Figure 2 of Erkut et al. (2016) comprised the characteristics of individual source data such as $\dot{M}$ and $B$ discussed by Erkut \& \c{C}atmabacak (2017). In other words, the correlation for the ensemble is a result of the superposition of the regions of model function fit to individual source data.

As shown in Equations (8)-(13), $\nu$ should be connected with $\dot{M}/B_{\ast}^2$ in the modulation of $B_{\ast}$. We obtain the solutions of 4U 0614+09, 4U 1608-52, 4U 1636-53 from our MHD model in different $B_{\ast}$ and select several groups of data with different $B_{\ast}$, which are shown in Figure 4. In the left panels of Figure 4, the twin kHz QPOs are arranged approximately in parallel for different $B_{\ast}$ from the balance condition at the magnetosphere radius. In the right panels of Figure 4, we obtain the relation between $\nu_{\rm lc}, \nu_{\rm uc}$ and $\dot{M}/B_{\ast}^2$ in the three sources. All the data of kHz QPOs in the three sources for different $B_{\ast}$ nearly converge into a group of curves, which means that $\nu_{\rm lc}, \nu_{\rm uc}$ and $\dot{M}/B_{\ast}^2$
 exhibit an approximative one-to-one relation in the range of a given $B_{\ast}$.

\begin{center}
\begin{figure*}[!htbp]
\label{fig4}
 \includegraphics[width=1.0\columnwidth]{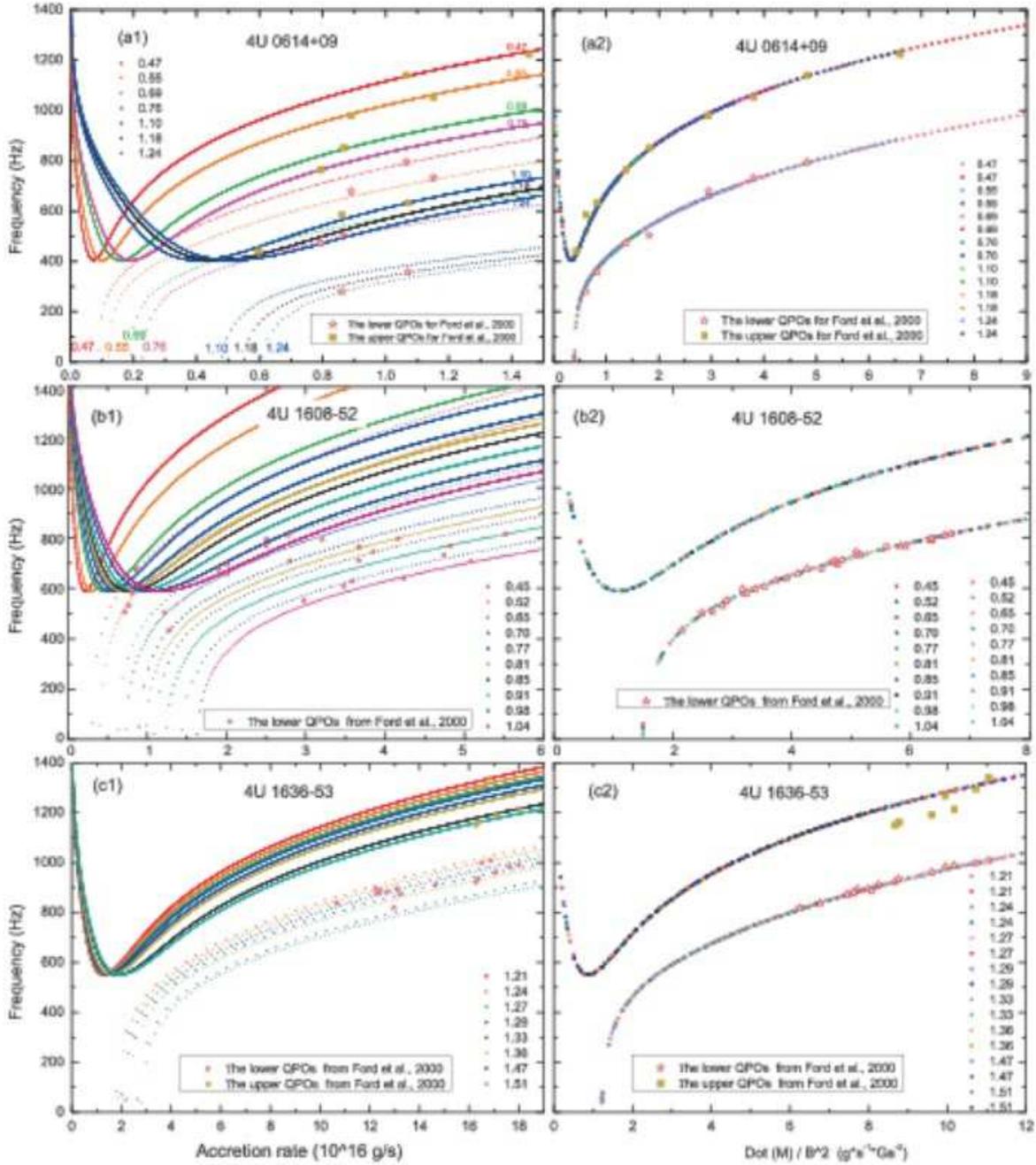}
 \caption{Relations between $\nu_{\rm lc}, \nu_{\rm uc}$ and the accretion-related parameters ({\it left}: accretion rate, {\it right}: $\dot{M}/B_{\ast}^2$) with different magnetic field ($0.47,0.55,0.69,0.76,1.10,1.18,1.24 \times 10^8\ {\rm G}$ for 4U 0614+09; $0.45,0.52,0.65,0.70,0.77,0.81,0.85,0.91,0.98,1.04 \times 10^8\ {\rm G}$ for 4U 1608-52; $1.21,1.24,1.27,1.29,1.33,1.36,1.47,1.51 \times 10^8\ {\rm G}$ for 4U 1636-53) (for the measured data: Ford et al., 2000. In the right panels, $B_{\ast}$ for Ford et al. is selected according to the left panel by eyes).}
\end{figure*}
\end{center}

\begin{center}
\begin{figure*}[!htbp]
\label{fig5}
 \includegraphics[width=0.5\columnwidth]{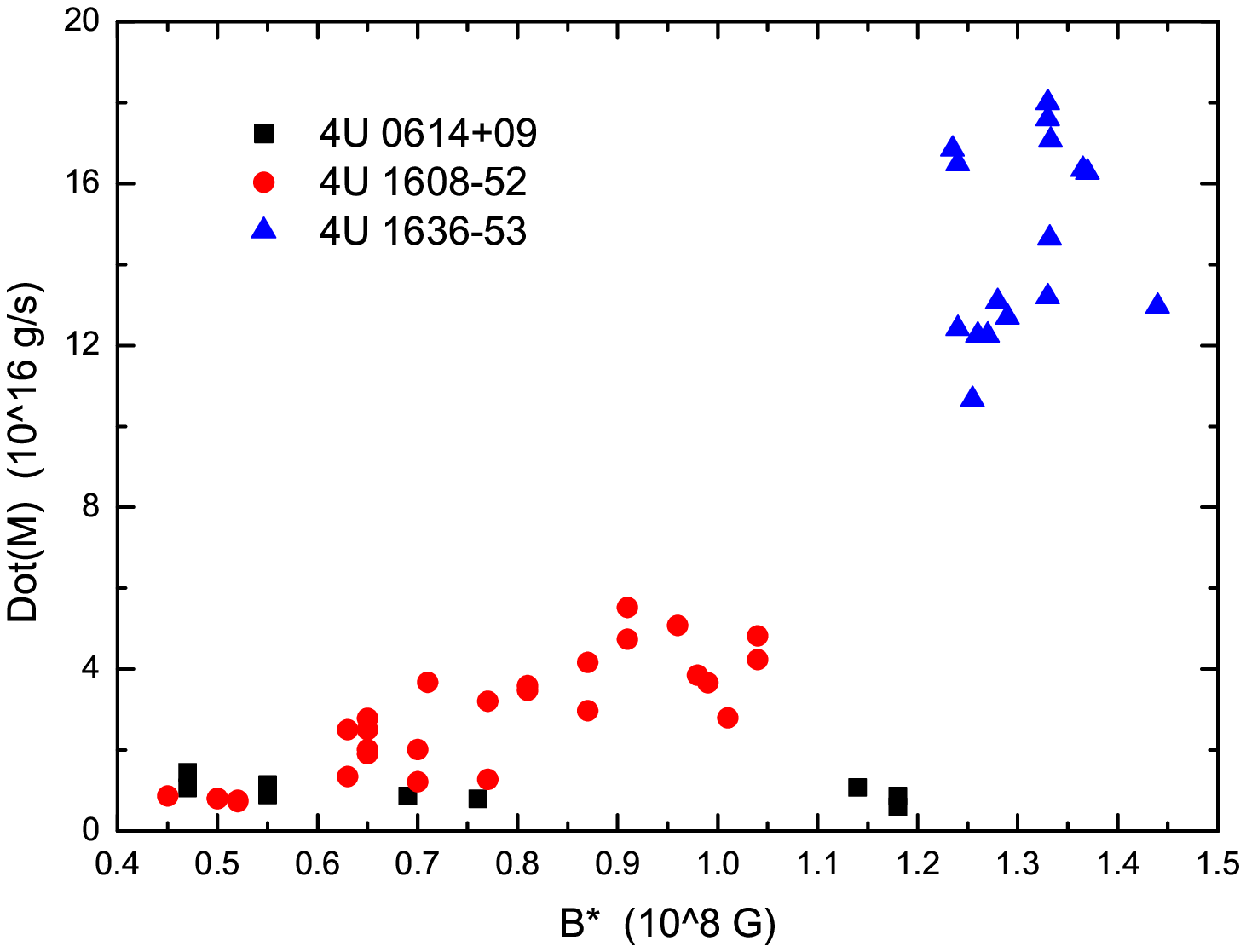}
 \caption{Relations between $B_{\ast}$ and  $\dot{M}$ in the three sources: 4U 0614+09, 4U 1606-52, 4U 1636-53 estimated from the measured data of Ford et al., 2000.}
\end{figure*}
\end{center}

As seen from the left panel of Figure 4, a stable magnetic field lead to a stationary relation between $\nu_{\rm lc}, \nu_{\rm uc}$ and $\dot{M}$, but a variable magnetic field does not. As shown in Figure 5, there is no correlation between $\dot{M}$ and $B_{\ast}$ in 4U 0614+09 and 1636$-$53, and there is only a weak correlation between $\dot{M}$ and $B_{\ast}$ in 4U 1608$-$52. In other word, the one-to-one correlation in each source between the QPO frequency and $\dot{M}/B_{\ast}^2$ shown in the right panels of Figure 4, is determined by the combination of the independent parameters $\dot{M}$ and $B_{\ast}$. The possible origins of the variations of the effective magnetic field will be discussed in section 4.
Therefore the twin kHz QPOs in one source can be described well by the relation between $\nu_{\rm lc}, \nu_{\rm uc}$ and $\dot{M}/B_{\ast}^2$. Similar to van der Klis (2001), the factor $B_{\ast}^2$ is mathematically equivalent to the composite luminosity.

Because the very long expressions of the analytic solutions for Equation (8) are too complex to be used, we only obtained the numerical solutions.
According to the one-to-one relation, we fit our numerical solutions of Equation (8) with several power-law functions expressed by the lines in Figure 6. The fitting models and their related parameters are listed in Table
2. As seen in Table 2, the the square of correlation coefficent ($C_{\rm r}^2$) of the fits for the relations between $\nu_{\rm uc}, \nu_{\rm lc}, \nu_{\rm sc}$ and $\dot{M}$ are near ``1'', which means a very strong correlation. We fit ${\nu_{\rm s}}$ with the power-law model (${\nu_{\rm s}}=a\ast(x-b)^\beta$) in order to keep the consistency of the model, but the reduced $\chi^2$ of the fits are much larger and their $C_{\rm r}^2$ are slightly smaller than those with the model ($\nu_{\rm s}=a\ast(b-x)^\beta+c_{2}$), so it is discarded.
The selected fitting model for ${\nu_{\rm s}}$ in Table 2 is not consistent with that of $\nu_{\rm u}$, $\nu_{\rm l}$ and the different places generated kHz QPOs in the accretion disc may lead to that.

 As seen in Table 2 and Figure 6, $b$
is close to $\dot{M}/B^2$ in the transition radius for the lower QPOs
 , i.e. $b\simeq\frac{\dot{M}}{B^2}\mid_{r=r_{\rm t}}$ where $k^2 v_{\rm A}^2+\omega_{\rm k}^2-\Omega^2 = 0$.
Finally, the relation between $\dot{M}$ and $\nu_{\rm u}$, $\nu_{\rm l}$ can be fitted by a power-law function, which can be expressed as follows,
\begin{equation}
\label{eq14}
\nu=a*(\frac{\dot{M}}{B^2}- b)^\beta.
\end{equation}

\begin{center}
\begin{figure*}[!hbp]
\label{fig6}
\centering
 \includegraphics[width=0.5\columnwidth]{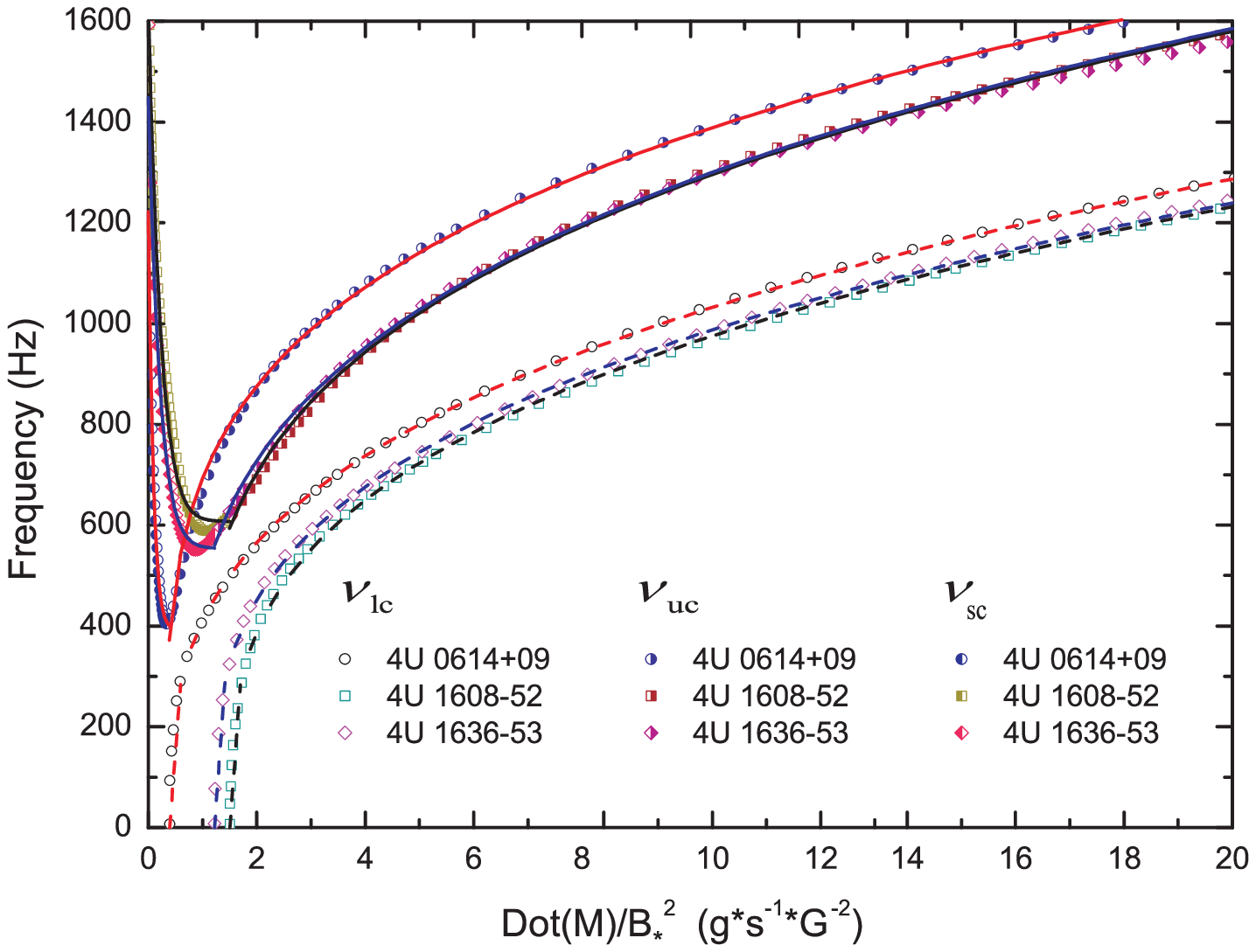}
\caption{ Relations between $\nu_{\rm lc}, \nu_{\rm uc}$ and $\dot{M}/B_{\ast}^2$ in our MHD model and fitting by several power-law functions in 4U 0614+09, 4U 1608$-$52, 4U 1636$-$53. The upper solid lines (the branches on the left of $r_{\rm t}$: for $\nu_{\rm sc}$; the branches on the right of $r_{\rm t}$: for $\nu_{\rm uc}$) express the result fitted the upper solutions of Equation (6). $\nu_{\rm lc}$ is fitted by the dashed lines.}
\end{figure*}
\end{center}

\begin{table*}[!htbp]\footnotesize
 \centering
  \caption{This table lists the models and parameters used in our fitting for the twin kHz QPOs ($\nu_{\rm u}$, $\nu_{\rm l}$) and the single kHz QPOs ($\nu_{\rm s}$): $x=\dot{M}/B_{\ast}^2$. }
  \label{t:2}
 \scriptsize{
 \setlength{\tabcolsep}{1.5pt}
  \begin{tabular}{@{}lcccccccc@{}}
   \hline
  \hline
{sources} & branches & model& $a$ & $b$& $\beta$& $c_{2}$ &Reduced $\chi^2$& $C_{\rm r}^2$\\
    \hline
 {4U 0614+09} &  $\nu_{\rm u}$ & ${\nu_{\rm u}}=a\ast(x-b)^\beta$ & $765.978\pm0.470$ & $0.346\pm0.001$ & $0.256\pm0.001$ & $\sim$ &117.910 &0.9999\\
   &  $\nu_{\rm l}$ & ${\nu_{\rm l}}=a\ast(x-b)^\beta$ &$484.272\pm0.127$ & $ 0.397\pm0.001$ &$0.328\pm0.001$ & $\sim$ &16.022 & 0.9992\\ & $\nu_{\rm s}$ & $\nu_{\rm s}=a\ast(b-x)^\beta+c_{2}$ & $16.085\pm192.547$  & $1.332\pm0.896$ & $13.763\pm9.500$  & $393.709\pm8.536$&986.639&0.9796\\
  \hline
 {4U 1608$-$52} &  $\nu_{\rm u}$ & ${\nu_{\rm u}}=a\ast(x-b)^\beta$ & $683.228\pm0.863$ & $0.890\pm0.005$ & $0.284\pm0.001$ & $\sim$&122.481&0.9992\\
   &  $\nu_{\rm l}$ & ${\nu_{\rm l}}=a\ast(x-b)^\beta$ &$485.283\pm0.317$ & $1.507\pm0.001$ &$0.319\pm0.001$ & $\sim$ &112.091&0.9991\\ & $\nu_{\rm s}$ & $\nu_{\rm s}=a\ast(b-x)^\beta+c_{2}$ & $0.002\pm0.019$  & $2.918\pm1.051$ & $12.238\pm4.612$  & $608.371\pm3.904$ &3261.981&0.9742\\
  \hline

{4U 1636$-$53} &  $\nu_{\rm u}$ & ${\nu_{\rm u}}=a\ast(x-b)^\beta$ & $673.704\pm0.711$ & $0.692\pm0.004$ & $0.289\pm0.001$ & $\sim$&35.070&0.9992\\
   &  $\nu_{\rm l}$ & ${\nu_{\rm l}}=a\ast(x-b)^\beta$ &$488.279\pm0.264$ & $1.219\pm0.001$ &$0.317\pm0.001$ & $\sim$ &38.886&0.9993\\ & $\nu_{\rm s}$ & $\nu_{\rm s}=a\ast(b-x)^\beta+c_{2}$ & $(3.123\pm51.588)\ast10^{-5}$  & $3.211\pm1.623$ & $14.719\pm7.793$  & $553.921\pm3.634$&425.576&0.9908\\
  \hline
\end{tabular}}
\end{table*}

\subsection{The NS Parameter Dependence}

As seen from Equations (9) to (14), the results of Equation (6) depend on not only $\dot{M}$ and $B_{\ast}$ but also $M,\ R$ and spin frequency of the NS ($\nu_{\rm NS}$) in a LMXB. Erkut et al. (2016) also studied systematically the effect of different masses and radii on kHz QPO frequencies in the the boundary region model of kHz QPOs. We solve Equation (6) with different $M,\ R,\ \nu_{\rm NS}$ again and the results are shown in Figure 7. In the calculation, we consider the dependence of $\nu_{\rm u}$ and $\nu_{\rm l}$ on one parameter while keeping other parameters their typical values.

\begin{center}
\begin{figure*}[!htpb]
\label{fig7}
\centering
 \includegraphics[width=0.6\columnwidth]{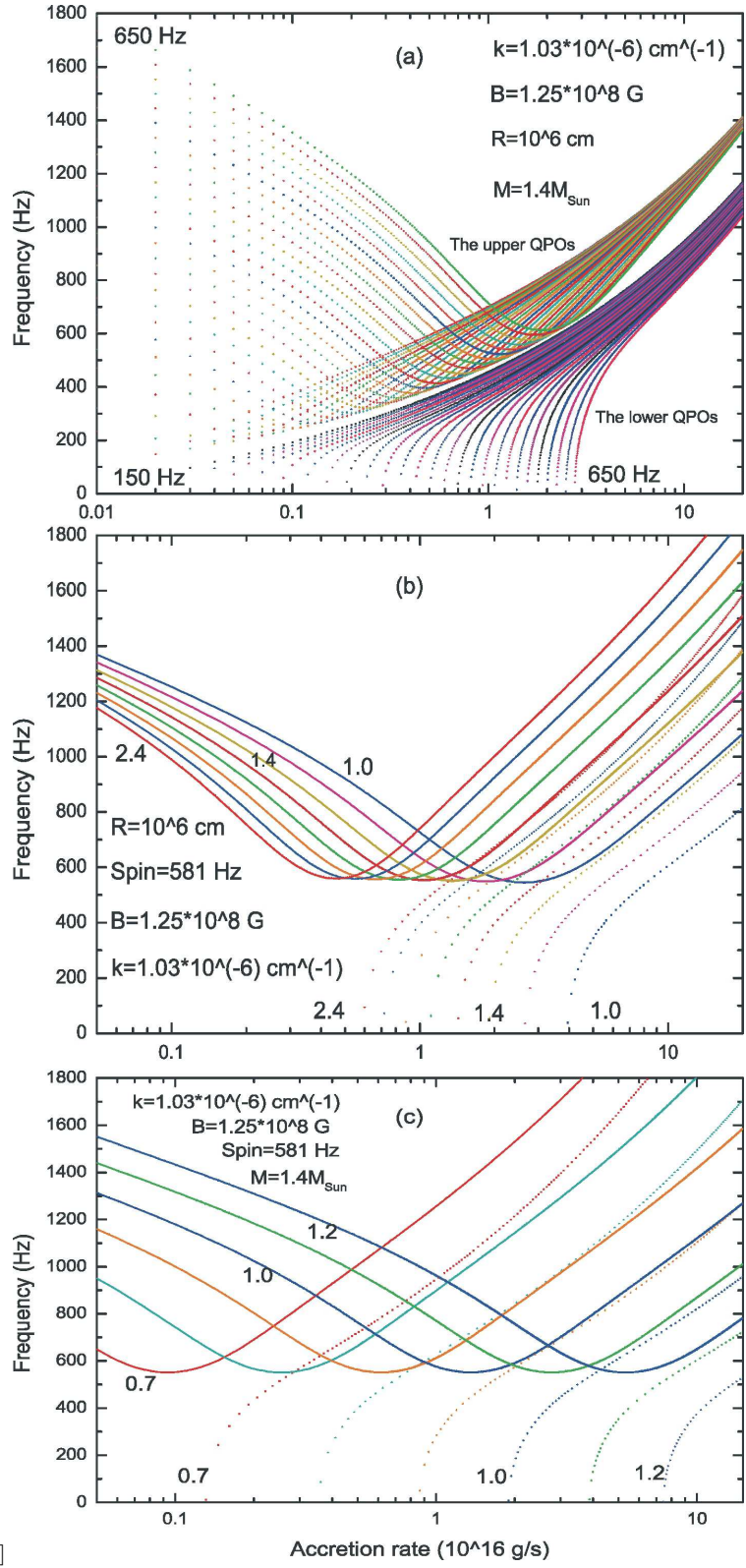}
 \caption{Relation between $\nu_{\rm lc}, \nu_{\rm uc}$ and $\dot{M}$ for different parameters of the NS (top panel: $\nu_{\rm NS}$; middle panel: $M$; bottom panel: $R$). }
\end{figure*}
\end{center}

As shown in the top panel of Figure 7, the line on the relation between $\nu_{\rm u}, \nu_{\rm l}$ and $\dot{M}$ shifts from left to right with increasing $\nu_{\rm NS}$. From left to right, $\nu_{\rm NS}$ increases with a $10\ {\rm Hz}$ spacing. The upper dotted lines express  $\nu_{\rm us}$ and the lower dotted lines express  $\nu_{\rm ls}$. The frequencies of the single kHz QPOs decrease from 1700 Hz to 100 Hz when the spin is decreased at a low accretion rate. The frequencies of the twin kHz QPOs, especially for $\nu_{\rm us}$, have smaller variation range for a high accretion rate than for a low accretion rate.
With increasing $\nu_{\rm NS}$, the characteristic frequencies of kHz QPOs in NS-LMXBs decrease at the same accretion rate, which might reflect the increase of the inertial centrifugal force.

In the middle panel of Figure 7, we obtain the relation between $\nu$ and $\dot{M}$ with increasing $M$. Contrary to the changing trend of lines in the top panel of Figure 7, the lines in the middle panel shift to the left when $M$ increases from $1.0 M_{\odot}$ to $2.4 M_{\odot}$ with a $0.2M_{\odot}$ spacing, i.e. $\nu_{\rm us}, \nu_{\rm ls}$ increase with increasing $M$ for the same $\dot{M}$, which means the effect from the increasing gravity.

In the bottom panel of Figure 7, the dependence of the kHz QPOs on $R$ is shown. The lines shift to the right while the radius of a NS increases from $0.7\times10^6\ {\rm cm}$ to $1.2\times10^6\ {\rm cm}$ with a $10^6\ {\rm cm}$ spacing. The twin kHz QPOs decrease fast with the increasing $R$ at the same $\dot{M}$, which means the radius will exert a tremendous influence on the kHz QPOs due to $\omega_{\rm k} \propto R_{6}^{-9/4}$.

From Figure 7, we can find that the similar range of kHz QPO frequencies (more than two orders of magnitude) is observed in different sources with different $\dot{M}$ if the adopted parameters are appropriate. That conclusion is coincident with the observations that the similar range of QPO frequencies is observed in different sources with different X-ray luminosity (spreading more than 2 orders of magnitude) (van der Klis 2001; Ford et al. 2000). As described by van der Klis (2006), $\nu$ may depend not only on $L_{\rm x}$, but also on other parameters, which leads to that different sources have different $\nu-L_{\rm x}$ tracks. In this study, the spin, mass and radius of a NS and $B_{\ast}$ lead to different tracks in different sources.

\section{DISCUSSION}

The MHD wave originates from a small disturbance, which may come from some instabilities. A lot of instabilities such as gravitational instability (Lin \& Pringle 1987), magnetorotational instability (MRI; Balbus \& Hawley 1998) and Rossby wave instability (Lovelace et al. 1999) can lead to the evolution of disks and creating complexity (Fung \& Artymowicz 2014). The interaction at the inner edge of the disc is a complicated process that depends on the field strength and disc properties such as viscosity and magnetic diffusivity. In our ideal model, the vacuum electroconductibility is supposed as an infinite quantum, which means that the magnetic diffusivity is neglected. However, magnetic diffusivity can be one origin for the deformation of a NS magnetic field and an instability. In addition, the viscosity as a constant ($\alpha$) is substituted into this model, but $\alpha$ may be dependent on intrinsic disc properties. Potter and Balbus (2014) found cyclic flaring in the disc when $\alpha$ as a variable satisfied their instability criterion. That means the viscosity also may bring about a small disturbance and lead to MHD waves. A limited vacuum electroconductibility and a variable $\alpha$ viscosity are the main characteristics of the real accretion plasma by comparing with the plasma in our ideal MHD model. So a group of nonideal MHD equations may solve the problems about the origin of the MHD waves.

As seen from figure 4, the derived effective maximum magnetic field $B_{\ast}$ is about 2.6 times of the minimum $B_{\ast}$ for the NS in 4U 0614+09. That variable number is the same order of magnitude shown in the other two sources. However, the evolution of the interior magnetic field of a NS is very slow (Goldreich \& Reisenegger 1992; Xie \& Zhang 2011), i.e. $B_{\ast}$ can not be the surface magnetic field of a NS. Can the external magnetic field change so much? Guglielmino et al. (2010) showed that the magnetic field in the solar atmosphere increases to
four times by the multiwavelength observations of small-scale magnetic reconnection events. That means the magnetic field may change a lot in the event of magnetic reconnection. Rembiasz et al. (2016) also find that the magnetorotational instability can amplify initially the magnetic fields of a proto-NS to 10 times by the three-dimensional shearing-disc and shearing-box simulations of a region close to the surface of a NS. Patruno (2012) also found that the effective surface magnetic field of the accretion X-ray pulsar (HETE J1900.1$-$2455) decreased from $5\times10^8\ \rm G$ to $7\times10^5\ \rm G$ during the about 100 days decay.

In our model the form factor ($c$) varies, which leads to the shift of different tracks. The change of $c$ originates from the change of the effective magnetic moment, or the effective surface magnetic field, i.e. the change of the magnetic field outside a NS. There are many factors leading to the change of the external magnetic field of a NS, which have been discussed by many authors. Two types of them including transient change and instantaneous change are related to change of the effective surface magnetic field.

%The secular change of the external magnetic field mainly comes from some instabilities, such as the magnetorotational instability (Guilet \& M$\ddot{\rm u}$ller 2015; Sai et al. 2013) and r-mode instability(Cuofano et al. 2013). Both instabilities affect the magnetic field in accretion disk but that r-mode instability mainly generates the strong magnetic field of the inner core of a NS (Cuofano et al. 2013).
%The nutation of the accretion disc may lead to the change of the misalignment angle between the rotational axis of an accretion disc and the spin axis of a NS, and hence the change of the boundary between the magnetosphere and the disc (i.e. $r_{\rm m}$).

The transient change of the external magnetic field in an accretion disc is produced by two factors, i.e. MRI and the screened effect of accreted plasma.
MRI leads to turbulence within a disc and drives an accretion due to the efficient outward angular momentum transport in a differentially rotating disc (e.g. Balbus \& Hawley 1991, 1992). So MRI may be one type of origin of the MHD wave and leads to the change of the magnetic field during tens of orbits (Balbus \& Hawley 1992; Guilet \& M$\ddot{\rm u}$ller 2015; Sai et al. 2013).
As described above, the effective surface magnetic field of HETE J1900.1$-$2455 was found to have changed by three orders of magnitude during about 100 days. Patruno (2012) considered that the external magnetic field was screened by freshly accreted plasma in burial models (Cumming et al. 2001; Cumming 2008). Cumming (2008) found that screening could be effective at the accretion rates of most steadily accreting neutron stars in LMXBs in our Galaxy. The magnetosphere radius can be rewritten as $r_{\rm m}\propto c_{\rm 3}B_{\rm NS} ^{-2/5}$, where $c_{\rm 3}$ is the factor due to the combination of the transient change of the external magnetic field originating from the MRI effect and the screening effect, which modulates the instantaneous change of the external magnetic field discussed below.

 %So the last magnetosphere radius can be expressed as, $r_{\rm m}\propto c_{\rm 3}B_{\rm NS} ^{-2/5}$, where $c_{\rm 3}$ is the nutation factor related to the misalignment angle.
The instantaneous change of the external magnetic field may be produced by two factors, i.e. the compression of accretion matter and the effect of reconnection or re-emerge of magnetic field lines. As described by Kulkarni \& Romanova (2013), the magnetic field is compressed by the accretion matter, when it resists and counter-balances the compression (Ghosh \& Lamb 1979). So the deformation of the magnetic field (e.g. from the impact of the accretion matter, Rezania \& Samson 2005; Kulkarni \& Romanova 2013; from the distortion of magnetic field lines in the differential rotation) will emerge, i.e. the magnetic field is no longer dipolar (Arons 1993).
The magnetosphere radius can be rewritten as $r_{\rm m}\propto c_{\rm 4}B_{\rm NS} ^{-2/5}$, where $c_{\rm 4}$ is the deformation factor that should be related to $\dot{M}$  and the magnetic pitch angel from the twist of the magnetic field lines
when the deformed magnetic field is considered. In our model, the MHD wave generated from some disturbance is considered as the origin of QPOs and different accretion rates lead to the different kHz QPOs in a track with the same effective magnetic field. However, the variation of $c_{\rm 4}$ might make the real track fluctuate around the track predicted in our MHD model.

Instantaneous change of the effective magnetic field of a NS may happen in a sudden outburst. Some repeated mini-outbursts with a timescale $\sim 10^5\ \rm s$, which were explained by disk instability, are frequently observed in some LMXBs such as 4U 1636-53 and 4U 1608-52 (Asai 2015, Belloni et al 2007). In addition, thermonuclear bursts (Ayasli  1982) also happen frequently. Linares (2012) discovered that thermonuclear bursts in IGR J17480$-$2446 gradually evolved into a QPO when the accretion rate increased, and vice versa. Patruno (2012) found that the magnetic field can re-emerge and return to an initial value of
$\sim 10^8\ \rm G$ in case that an outburst is over.
Simulations about MRI also show a cycle which corresponds to the rapid amplification of the magnetic field followed by decay through magnetic reconnections in LMXBs (Sano \& Inutsuka, 2001; Latter \& Papaloizou, 2012).
Reconnection or re-emerge of magnetic field lines from those instabilities and bursts may turn on a new cycle (accreting matter, screening, distorting and compressing the magnetic field lines, re-emerging or reconnecting of the magnetic field lines, accreting matter again). Now we bring the two factors
($c_{\rm 3}$ \& $c_{\rm 4}$) into a single factor ($c$), who is supposed as the form factor in Equation (8). Different tracks of kHz QPOs in a LMXBs can be explained as a compositive result on the transient change and the instantaneous change of the magnetic field.
\begin{center}
\begin{figure*}[!htpb]
\label{fig8}
\centering
 \includegraphics[width=1.0\columnwidth]{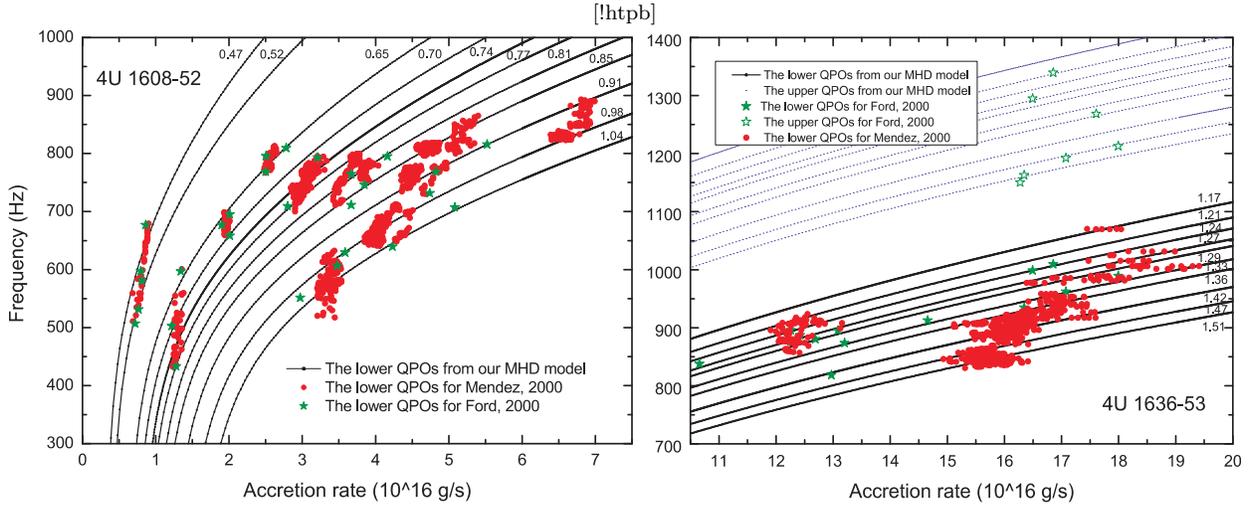}
 \caption{Relation between the frequencies of kHz QPOs and $\dot{M}$ deduced from the X-ray luminosity of the source ($L_{\rm X}=GM\dot{M}/R$) for a NS of $M = 1.4 M_{\odot}$ and $R=10 {\rm km}$. That X-ray luminosity is estimated through a comparison of the frequency vs. count rate data (M\'endez 2000) with the frequency vs. $L_{\rm X}$ data in Ford et al. (2000) (left panel for 4U 1608$-$52 and right panel for 4U 1636$-$53.}
\end{figure*}
\end{center}

The detailed relation between the frequencies of kHz QPOs and the estimated $\dot{M}$, which is deduced from the X-ray luminosity of the source ($L_{\rm X}=GM\dot{M}/R$) for a NS of $M = 1.4 M_{\odot}$ and $R=10\ {\rm km}$, is also shown in Figure 8, where luminosity is estimated through a comparison of the frequency vs. count rate data (M\'endez 2000) with the frequency vs. $L_{\rm X}$ data in Ford et al. (2000). Zhang et al. (1998) found that the correlation between the QPO centroid frequency and the measured flux in the 2-10 keV band is very similar to that between the frequency and the count rate. In Figure 8, $\dot{M}$ is roughly estimated by the hypothesis that count rate is proportional to the flux and thus $\dot{M}$ (e.g. Jahoda et al. 2006, Erkut \& \c{C}atmabacak 2017).
As seen from Figure 8, the systematic decrease in the slopes of parallel tracks and
tendency to cover higher frequency ranges as luminosity increases is also shown in the theoretical curves. However, the slight difference between the data in 4U 1608$-$52 and 4U 1636$-$ 53 and our MHD model may be due to two reasons. One is that the detailed energy spectrum is not considered, and the other is that $c_{\rm 4}$ is not a constant for different $\dot{M}$, i.e. the accretion rate leads to the small instantaneous change of $B_{*}$. If that is the key reason, our model predicts that the kHz QPOs will gradually migrate from one curve to another, which needs to be explored further.

Now we summarize the explanation of the ``parallel tracks" as follows.
\begin{enumerate}
\item Due possibly to, MRI, the screening effect of accreted plasma and some deformation of the magnetic field of the NS in LMXBs, the magnetic field near the surface of a NS ($B_{\ast}$) will be changed and is supposed to change from one characteristic magnetic field ($B_{\rm 1}$) to another ($B_{\rm 2}$), where the subscripts ``1'', ``2'' express $B_{\ast}$ before and after the change, respectively. Meanwhile, the shape of the magnetic field also changes.

\item A magnetosphere radius is determined by the balance of accretion plasma at the innermost boundary of an accretion disc (e.g. ${\frac{B^2}{8 \pi} \simeq P + \rho u^2}$, Eslner \& lamb 1977), thus it can be expressed as a function ($r_{\rm m}=f_{\rm1}(\dot{M},B_{\ast},M,R)$). Then the change of $B_{\ast}$ may lead to a change of $r_{\rm m}$.

\item According to our MHD model, the frequencies of kHz QPOs are mainly determined by $r_{\rm m}$ and can be expressed as a function, $\nu=f_{\rm 2}(\dot{M},B_{\ast},M,R, \nu_{\rm NS})$ when the {\bf effective} magnetic field are determined. Thus we can obtain a
    track ($\nu=f_{\rm 2}(\dot{M},B_{\rm 1},M,R, \nu_{\rm NS})$) for $B_{\rm 1}$ and another track ($\nu=f_{\rm 2}(\dot{M},B_{\rm 2},M,R, \nu_{\rm NS})$) for $B_{\rm 2}$ in a frequency-accretion plane.

\item When a mini-outburst or other outbursts take place, the effective magnetic field can jump or turn on a new round of cycle.

\end{enumerate}
Each track in the frequency-luminosity for one source is generated from the variable instantaneous accretion rate. The ``parallel tracks'' of kHz QPOs can be explained by that the varying $B_{\ast}$ leads to the shift of the track, i.e. the deformation of the magnetic field of a NS, MRI, the screening of plasma accreted onto the surface of a NS and some X-ray outbursts lead to the ``parallel tracks''.

In the Sun, there are three main groups of mechanisms invoked to explain the observed QPOs, i.e. an external modulation of the magnetic reconnection, an inherent effect in the magnetic reconnection process, modulations of the flare particles by waves in the flaring loop. The interpretation on MHD waves in the flaring loops is the most popular and fits most of the observational features (Nakariakov \& Melnikov 2009; Van Doorsselaere et al. 2016). Observations also found the central frequencies of QPOs in the Sun depend on the density of plasma and ambient magnetic fields. However the density associate with the magnetosphere radius by the definition of the magnetosphere radius in our model, the frequencies of kHz QPOs in our MHD model are determined by $\dot{M}$, the derived $B_{\ast},\nu_{\rm NS},M,R$.

In our MHD model, Shi, Zhang \& Li (2014) supposed that the accreted matter is transported to the bipolar of the NS after it arrives at the magnetosphere radius. The accretion matter rush to the two poles and the gravitational potential energy is mainly released there. Here we suppose that the radiative effect is weaker than the magnetic pressure outside the ISCO, and is neglected. The detailed result including the radiative effect will be explored in the future.

\section{CONCLUSION}
In this study we recalculated the modes of the MHD waves in the MHD model (Shi, Zhang \& Li 2014) of the kHz QPOs in NS-LMXB. We expanded the method to determine the wave number and magnetic field. We found that the change of the effective magnetic field $B_{\ast}$ from the balance at the magnetosphere radius hardly affects $\sigma$. After comparing the solutions of Equation (6) with observations, we obtained a good relation between the frequencies of the twin kHz QPOs and the accretion rates.
In the explanation, we obtained a simple power-law relation between the kHz QPO frequencies and an accretion-related parameters in those sources. Finally we studied the dependence of kHz QPO frequencies on the spin, the mass and the radius of a neutron star.
Our main results are summarized as follows.
\begin{enumerate}
\item The wave number in our MHD model of kHz QPOs can be determined by the the minimum deviation ($\sigma = \frac{1}{n} \sqrt{\Sigma (\nu_{\rm uc} - \nu_{\rm uo})^2}$).

\item The parameter ($\sigma$) can be thought as an invariant with changing $B_{\ast}$.

\item The ``parallel tracks'' of kHz QPOs can be explained by that the slowly varying $B_{\ast}$ leads to the shift of the track, which is generated from the variable instantaneous accretion rate.

\item The tracks for different $B_{\ast}$ from all the data of kHz QPOs
 converge into a group of curves, i.e. the frequencies of the twin kHz QPOs exhibit a one-to-one relation with $\dot{M} /B_{\ast}^2$ in the range of a given normal $B_{\ast}$.

 \item The relation between accretion rates ($\dot{M}$) and the frequencies of the twin kHz QPOs ($\nu_{\rm u},\ \nu_{\rm l}$) can be fitted by a power-law function ($\nu=a*(\frac{\dot{M}}{B_{\ast}^2}- b)^\beta$).

 \item The mass, radius and spin of the NS in a NS-LMXB will exert tremendous influences of kHz QPOs in our MHD model.
\end{enumerate}

\section*{acknowledgements}
The authors would like to acknowledge the anonymous referee for useful comments.
This work is supported by the National Natural Science Foundation
of China under grant Nos. 11563003, 11203009, 11133001, 11333004, 11373036 and 11133002, the National Key Research and Development Program of China (2016YFA0400803), and Key Research Program of Frontier Sciences, CAS, Grant NO. QYZDY-SSW-SLH008.

\clearpage

\label{lastpage}

\end{document}